\documentclass[10pt, reprint, aps, amsmath,amssymb,twocolumn,article]{revtex4-1}

\usepackage{epigraph}
\usepackage{graphicx}
\usepackage{dcolumn}
\usepackage{algorithm,algpseudocode}
\usepackage{bm}
\usepackage[font=scriptsize]{caption}
\usepackage{booktabs}
\usepackage{subcaption,graphicx}
\algnewcommand{\To}{\textbf{To }}
\algnewcommand\Input{\item[\textbf{Input:}]}%
\algnewcommand\Output{\item[\textbf{Output:}]}%

\begin{document}

\title{An Early Modeling Approach to Digital Electronics}

\author{Luciano da F. Costa}
\email{luciano@ifsc.usp.br, copyright LdaFcosta}
\affiliation{S\~ao Carlos Institute of Physics, IFSC-USP,  S\~ao~Carlos, SP,~Brazil}

\date{\today}

\begin{abstract}
An Early modeling approach of transistors characterized by simplicity and accuracy in representing intrinsic non-linearities
is applied to the characterization of propagation delay and level transition switching properties of NPN and PNP small signal
transistors.  Eight types of devices were considered, each represented by 5 samples taken from the same lot, totaling
20 NPN and 20 PNP transistors. Four switching time measurements were experimentally obtained, and the transistors also
had their Early parameters $V_a$ (the Early voltage) and $s$ (a proportionality constant such that $R_o = 1/tan(s I_B)$ 
accurately estimated
by using an experimental-numeric procedure that involves Hough transform accumulation in order to identify the crossing of the
base current ($I_B$) indexed characteristic isolines, yielding the respective $V_a$.  The timing measurements exhibited strong
positive Pearson correlations when taken pairwise.  When these measurements were compared individually to the respective 
Early parameters, no significant Pearson correlation was obtained.  However, a strong relationship was observed between 
the product of the two Early
parameters and the ratio between the fall and rise time.  A Pearson correlation coefficient of 0.78 was observed between these
variables in the case of NPN devices.  This suggests that transistors with larger average current gain tend to have
more similar rise and fall times.   The different relationship observed for PNP devices (Pearson 0.41) suggests some intrinsic
difference in the way the Early parameters influence the rise and fall times of small signal transistors.
\end{abstract}

\maketitle

\setlength{\epigraphwidth}{.49\textwidth}
\epigraph{``You may delay, but time will not.'}{B. Franklin}

\section{Introduction}
The most basic piece of information is the \emph{bit}.  Conceptualized by Claude Shannon in the 40s, the bit ended up
constituting the basis of almost every computer machine nowadays, including all digital systems.  In addition to its simplicity, 
digital representation and handling of information has the great advantage of being robust to noise. 

While \emph{linearity} represents the main objective of analog electronics, \emph{switching speed} provides one of 
the main interests in digital electronics (e.g.~\cite{roehr:1963,gray:1969,prasad:2009}).  As digital computing requires quick progress through 
a sequence of states, switching speed
becomes the main factor limiting computing velocity.   In general, it is the stray and intrinsic capacitances and inductances that limit 
switching speed by slowing down the digital transitions and implying in propagation delays.  These effects are governed by 
respective \emph{time constants} $\tau$ that, in the case of capacitance ($\tau = R C$) is proportional to the time required to 
charge/discharge the capacitors.  Interestingly, at this point digital electronics meets analog electronics, as these charging/discharging
are inherently analog effects.  Indeed, in a strict sense there is no purely digital electronics, as any state transition in the real-world
requires an \emph{analog} progression of values.

Transistors are the basis of modern electronics, both analog and digital.  Though often simplified as being linear, transistors are
actually highly non-linear devices.  Most of these linearities stem of the geometric structure exhibited by the characteristic isolines.
As a consequence of the Early effect (e.g.~\cite{early:1952,jaeger:1997}), these isolines cross one another, and with the $V_C$ (collector voltage) axis, at a 
well-defined point $(V_C = V_a, I_C = 0)$, where $V_a$ is the \emph{Early voltage}.  While the Early voltage has been considered
in electronics, usually as a means to conceptualize the output resistance variation exhibited by transistors, and also as part of the
Gummel-Poon model~\cite{gummel:1970,moinian:2000}, it is not enough, when taken alone, to provide a complete description of the 
characteristics of junction transistors.  

More recently, an approach has been reported that indicates that by incorporating a second, new proportionality parameter $s$,
it is possible to obtain a relatively accurate and yet simple model of transistors.  The main advantage of this approach derives from the
fact that neither $V_a$ or $s$ vary during the transistor operation.   This is not the case with other transistor parameterizations, such
as the traditional adoption of the current gain $\beta$ and output resistance $R_o$, both of which vary with both $V_C$ and $I_C$. 
Moreover, as the collector  output resistance has been experimentally shown~\cite{costaearly:2017,costaearly:2018,costaequiv:2018}
to vary as $R_o = 1/tan(s I_B)$, where $I_B$ is the base current, it becomes possible to derive a simple equivalent
circuit~\cite{costaequiv:2018} consisting of a fixed voltage source
$V_a$ in series with a varying output resistance $R_o = 1/tan(s I_B)$.  As $I_B$ is typically very small, it is often possible to use
the approximation $R_o \approx 1/(s I_B)$. Interestingly, the average current gain can be approximated
as $\beta \approx - s V_a$.  Observe that it was precisely the introduction~\cite{costaearly:2017,costaearly:2018} of the proportionality 
parameter $s$ that not only provided 
a direct indication of the output resistance $R_o(I_B)$, but also allowed the Early voltage to be related to the average
current gain $\langle \beta \rangle$.

This approach relying on the Early voltage $V_a$ and the complementary proportionality constant $s$ has been called the 
\emph{Early model} of small signal transistors~\cite{costaearly:2017,costaearly:2018,costaequiv:2018}.  
This approach has been successfully applied to a number of important problems in analog electronics.  First, it allowed several
real-world silicon and germanium junction small signal transistors, of types NPN and PNP, to be mapped into the 
$(V_a,s)$ space~\cite{costaearly:2018,germanium:2018,costaequiv:2018},
defining a prototype distribution of the four groups in the Early parameter space.  
This revealed that NPN types have smaller parameter variation, while being
characterized by larger $V_a$ magnitudes and smaller values of $s$.  However, the centers of mass of both NPN and PNP silicon
groups resulted very near the current gain isoline $\langle \beta \rangle = 240$, the same being observed for the germanium
NPN and PNP groups, but with $\langle \beta \rangle = 130$ in this case.  Silicon and germanium transistors tended to
``gravitate'' along two respective ``belts'' along the respective current gain isolines.  Interestingly, almost negligible overlap has been verified
between the NPN and PNP groups, be it regarding silicon or germanium devices.  These results show that real-world transistors
have intrinsic specific properties that influence their linearity, output resistance and, therefore, performance in amplification.

The above results motivated the proposal of an Early equivalent model~\cite{costaequiv:2018}, which consists of a fixed voltage
source $V_a$ in series with a variable output resistance $R_o = 1/tan(s I_B)$.   This equivalent circuit allowed the mathematical 
analysis of a simplified common emitter configuration~\cite{costaequiv:2018}, as well as the more complete configuration including 
negative feedback~\cite{commemitt:2018}.  In the latter case, it was possible to derive equations for the gains, leading to an accurate 
analysis of the effects of the $V_a$ and $s$ parameters
on current and voltage gains, as well as the implied non-linearities.  Other related developments include the analysis of capacitive
loads on amplification~\cite{costareact:2018}.

The efficacy of the Early approach to characterize and analyze analog circuits motivates its possible extension to digital electronics.
In particular, would it be possible to correlate the $V_a$ and $s$ parameters with propagation delays and switching transistors?
If so, the Early approach could be extended to the characterization of the switching properties of real-world transistors, paving
the way to several applications as well as analytical investigations.   This provides the main motivation for the present work.

We start by briefly revising the Early modeling approach, and then present some basic principles regarding the characterization
of switching properties.  Following, experimental measurements of the delay and transition switching times, and then of 
$V_a$ and $s$ parameters are presented and related one another.   Interesting results are then presented and discussed, and
the article concludes by outlining prospects for further related research.

\section{Brief Review of the Early Modeling Approach} \label{sec:Early}

\begin{figure}[h!]
\centering{
\includegraphics[width=8cm]{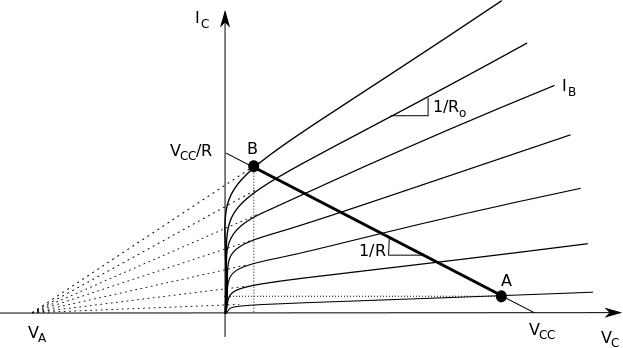}
\caption{The geometry underlying the Early modeling approach of the common emitter circuit in Figure~\ref{fig:commemitt}.}
\label{fig:early}}
\end{figure}

\begin{figure}[h!]
\centering{
\includegraphics[width=5cm]{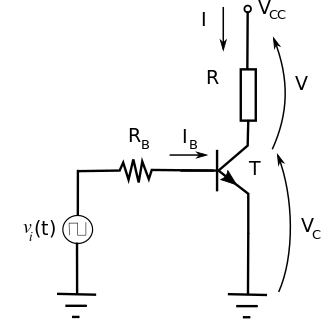}
\caption{The ``common emitter' circuit configuration considered in this work.}
\label{fig:commemitt}}
\end{figure}

Figure~\ref{fig:early} depicts the Early approach to modeling small signal transistors while taking into account the
respective non-linearities.  This refers to the ``common emitter'' circuit configuration shown in Figure~\ref{fig:commemitt}.
The $(V_C,I_C)$ space corresponds to the operation space of the transistor, which is restricted by the purely resistive
load $R$ to excursion along the load line passing through the points $(V_{CC},0)$ and $(0, V_{CC}/R)$.  A characteristic
isoline curve is defined for each value of the base current $I_B$, and these isolines intersect at the point $(V_a,0)$ 
along the $V_C$ axis, where $V_a$ is the \emph{Early voltage}.  The circuit excursion is, however, limited by the
cut-off (point $A$) and saturation (point $B$) regions of the transistor. The tangent of each isoline corresponds to the 
reciprocal of the collector output resistance $R_o(I_B)$.  

The Early modeling of the circuit in Figure~\ref{fig:commemitt} yields the following current and voltage equations:

\begin{eqnarray}
  V_C = I R_o(I_B) + V_a =  I/tan(s I_B) + V_a  \\
  I = I_C = V/R 
\end{eqnarray}

Observe that the collector output resistance is a function of $I_B$, i.e.~$R_o(I_B) = 1/tan(s I_B) \approx 1/(s I_B)$.
Thus, despite the extreme simplicity of these equations, they are still capable of incorporating, to a good level of accuracy,
the amplification non-linearity implied by the converging isolines.

In the case of the input mesh, we take the traditional approach $V_B = V_r + R_r I_B$, where $V_r$ is the fixed offset 
($\approx 0.6V$ for silicon) and $R_r$ is the input resistance.  

Given a transistor, its parameters $V_a$ and $s$ can be easily estimated through a experimental-numeric procedure~\cite{costaearly:2018}.
First, the characteristic isolines are acquired by making $V_{CC} = 0, \Delta V_C, 2 \Delta V_C, \ldots$ and measuring
$V_C$ for $I_B = 0, \Delta I_B, 2 \Delta I_B, \ldots$.  This can be conveniently achieved by using a microcontrolled
system~\cite{costaearly:2018}.  Once the isolines have been obtained and conditioned~\cite{costaearly:2018}, respective straight 
lines are estimated by using minimum squared linear regression.  The determination of the intersection of these lines can then be effectively accomplished  by using
a Hough transform accumulation scheme~\cite{costaearly:2018}, which allows eventually diverging isolines not to interfere with the
peak of the intersections.  The proportionality parameter $s$ can be easily obtained by linear regression of
the angle $\theta$ of the obtained isolines in terms of $I_B$.  This procedure has been 
verified~\cite{costaearly:2018,germanium:2018,costaequiv:2018}, for hundreds of small
signal silicon and germanium transistors, to yield an accurate and stable estimation of both $V_a$ and $s$.

It is interesting to recall that the above outlined Early modeling approach of transistors and related circuits derives part of
its simplicity from the experimentally verified~\cite{costaearly:2017,costaearly:2018} tendency of $\theta$ to vary linearly 
with $I_B$ through $s$, i.e. $\theta = s I_B$.  This approach can be adapted in cases where this dependence may 
follow a different relationship.

\section{Switching Measurements}

\begin{figure}[h!]
\centering{
\includegraphics[width=7.5cm]{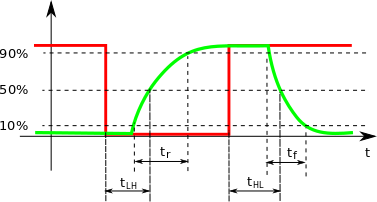}
\caption{The four switching measurements adopted in the present work, corresponding to:  the propagation delay $t_{LH}$
from low to high levels; the rise time $t_r$; the delay $t_{HL}$ from high to low levels; and the falling time $t_f$.}
\label{fig:waves}}
\end{figure}

There are at least two main types of switching effects to be considered in digital electronics: (i) propagation delays; and (ii) transition
times.   Figure~\ref{fig:waves} shows the input digital signal (in red) and the respective output (in green).  Four switching
measurements can be defined: (a) the propagation delay $t_{LH}$ from low to high levels; (b) the rise time $t_r$; (c) the
propagation delay $t_{HL}$ from high to low levels; and (d) the fall time $t_f$.

So, given a switching circuit such as that in Figure~\ref{fig:commemitt}, an input signal can be driven into it, and the 
above mentioned four measurements obtained by using a logging system or a digital oscilloscope.

\section{Experimental Set-Up}

Eight types of small signal silicon transistors (4 NPN and 4 PNP) were considered, each represented by 5 respective samples
taken from the same lot.  These devices are all in plastic package (TO-92) and correspond to models commonly used in discrete
electronics.  First, these transistors had their Early parameters $V_a$ and $s$ estimated by using the methodology summarized
in Section~\ref{sec:Early}.  Then, they were mounted in the circuit configuration shown in Figure~\ref{fig:commemitt}, having
$R_B = 100 k \Omega$, $R = 1 k \Omega$, and $V_{CC} = 8V$.  The input signal was a square wave derived from a signal 
generator, with $V_L = 0V$ and $V_H = 8V$.  Both the input and  the output signal $V_C$ were monitored with a digital oscilloscope,
so that the adopted four switching times could be estimated.  For simplicity's sake \emph{ all the currents and voltages of PNP transistors
are henceforth shown with inverse signs, allowing these devices to be discussed in the same way as the NPN counterparts}.

\section{Switching Time Results}

\begin{figure}[h!]
\centering{
\includegraphics[width=8.5cm]{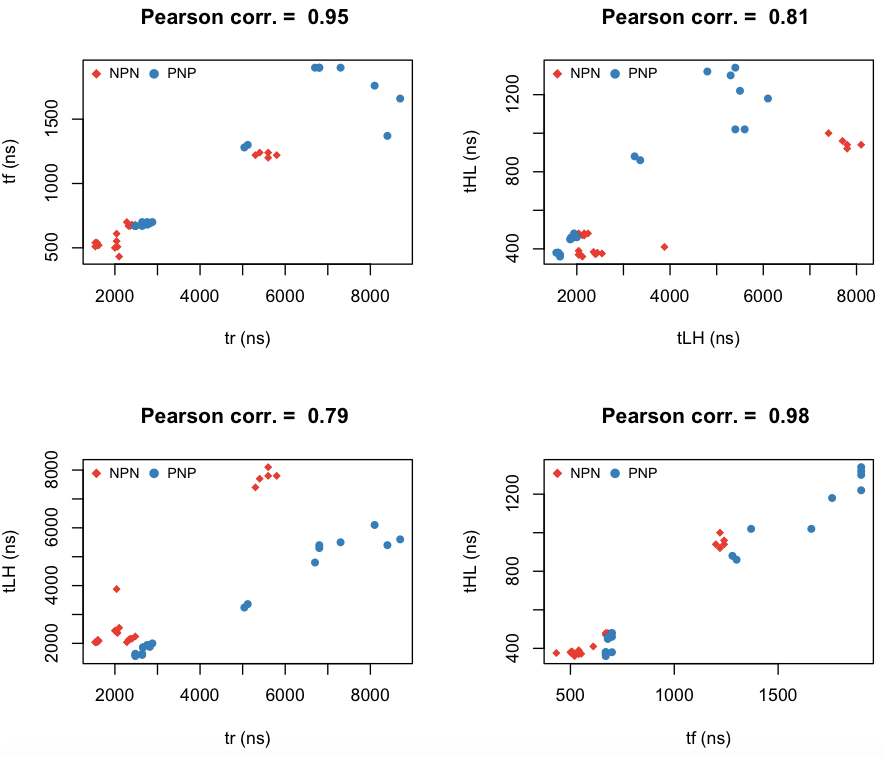}
\caption{Scatterplots and Pearson correlations obtained for: (a) $t_r \times t_f$; (b) $t_{LH} \times t_{HL}$; (c) $t_r \times t_{LH}$; 
and (d) $t_f \times t_{HL}$. High positive Pearson correlations are obtained in all cases, but the cases involving $t_f$ yielded the
highest Pearson correlations.}
\label{fig:times_correls}}
\end{figure}

Figure~\ref{fig:times_correls} shows the scatterplots obtained for (a) $t_r \times t_f$; (b) $t_{LH} \times t_{HL}$; (c) $t_r \times t_{LH}$; 
and (d) $t_f \times t_{HL}$.   Interestingly, all these cases yielded high positive Pearson correlations, with the two cases involving
$t_f$  -- i.e. (a) and (d) -- leading to particularly high correlation values.  These results indicate that the considered delay and transition
times are all interrelated.  However, observe that the positive transitions (i.e.~$t_r$ and $t_{LH}$) have markedly different values
from the negative transitions (i.e.~$t_f$ and $t_{HL}$).  Actually, the latter take place about 4 times faster than the former.  This suggests
that the adopted transistors in the considered circuit configuration can drain the stray capacitances much faster than charging them.  

The scatterplots in Figure~\ref{fig:times_correls} also identify the two types of considered transistors, i.e.~NPN and PNP.  
No special relationship can be identified in any of the 4 scatterplots regarding these two types.

\section{Early Parameters Estimation}

\begin{figure}[h!]
\centering{
\includegraphics[width=7cm]{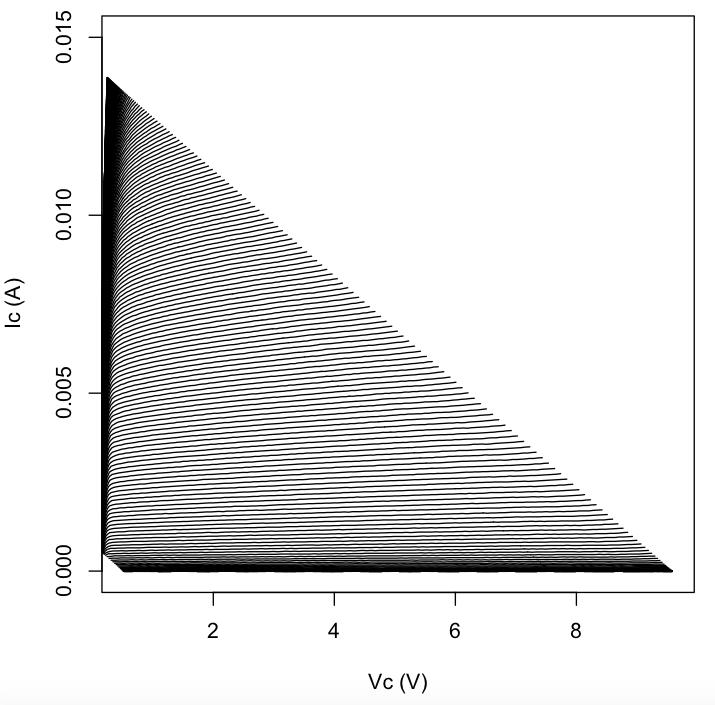}
\caption{The characteristic isolines typically obtained for one of the adopted PNP transistors.  This result is shown as received from the
acquisition system, without any noise treatment or signal conditioning.  Observe the fanning structure of the isolines, increasing
their slope along the bottom-up direction.  The saturation and cut-off regions can also be immediately identified.  The good 
quality of the data acquisition contributed to accurate estimation of the respective Early parameters.  In the case of the transistor
shown in this figure, we have $V_a = -35.00 V$ and $s = 5.95 V^{-1}$, yielding $\langle \beta \rangle \approx 208.25$.}
\label{fig:charact}}
\end{figure}

The characteristic isolines typically obtained for the considered transistors are illustrated in Figure~\ref{fig:charact}.  Great attention
to shielding and isolation of the digital and analog modules of the acquisition system~\cite{costaearly:2018} allowed high noise 
immunity and data
quality, which contributed to more accurate estimation of the Early parameters $V_a$ and $s$.   The fanning structure of the isolines
typically found in real-world transistors are evident in this figure.  They are the main source of transistor amplification non-linearity.
Observe the cut-off and saturation regions.  

\begin{figure}[h!]
\centering{
\includegraphics[width=9cm]{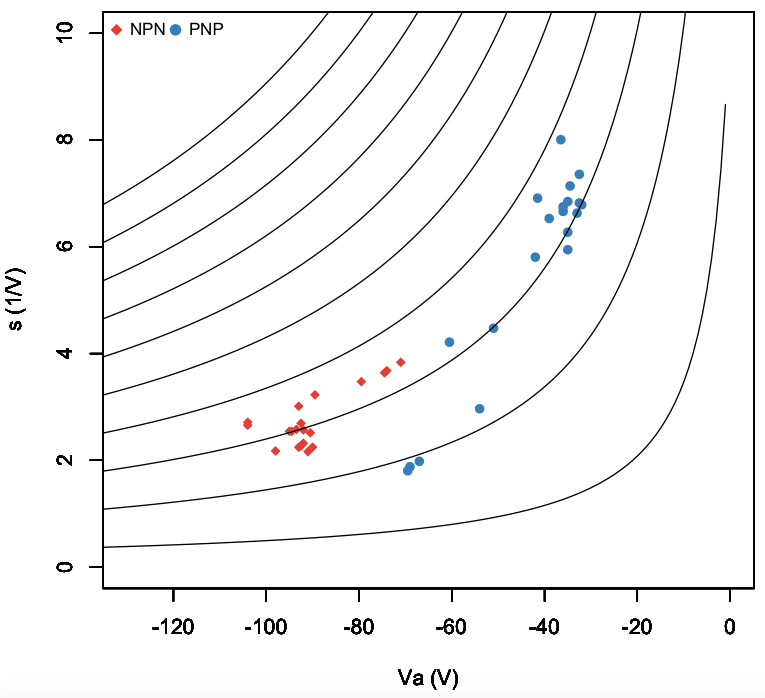}
\caption{The 40 considered transistors (20 NPN and 20 PNP) respectively mapped into the Early space $(V_a,s)$.
As this pair of parameters accurately represent the fanning isolines structures, the distribution of devices obtained in
the Early space is closely related to the respective electronic properties such as output resistance $R_o = 1/tan(s I_B)$
and current gain $\langle \beta \rangle \approx - s V_a$.  Observe that the NPN and PNP devices occupy mostly
non-overlapping respective regions, with the PNP transistors exhibiting larger values of $s$ and smaller values of $V_a$ 
magnitude.  The current gain isolines correspond (bottom-up) to $\langle \beta \rangle = 50, 150, 250, \ldots, 950$.
Most of the considered silicon transistors lie within the belt defined by $150 \leq \langle \beta \rangle \leq 350$.}
\label{fig:earlyspace}}
\end{figure}

The characteristic isolines obtained experimentally for each transistor were processed by the Early estimation procedure as
described above, so that each transistor could be characterized in terms of its respective parameter configuration $(V_a, s)$,
which are mapped into the henceforth called \emph{Early space}. Figure~\ref{fig:earlyspace} shows the distribution of all 40 transistors
(20 NPN and 20 PNP) in the Early space, showing also the average current gain isolines respectively to
$\langle \beta \rangle = 50, 150, 250, \ldots, 950$ (in bottom-up direction).

Several interesting features can be observed in Figure~\ref{fig:earlyspace}.  First, we have that, with a few exceptions observed for
PNP devices, most transistors resulted inside the current gain belt $150 \leq \langle \beta \rangle \leq 350$, with the isoline
$250$ being very near the center of mass of the two groups of devices.  This is in complete
agreement with previous studies~\cite{costaearly:2018,germanium:2018}, corroborating this as the region likely occupied by small signal transistors.  The PNP
transistors have smaller magnitudes of $V_a$ and higher values of $s$, exhibiting almost no overlap with the NPN counterparts.
This overall organization of NPN and PNP small signal transistors in the Early space has been called the \emph{prototypical space}
of transistors~\cite{costaearly:2018}.  The region  $ \beta \rangle \leq 150$ has been found to be populated by alloy junction germanium
transistors~\cite{germanium:2018}, the region of the Early space such that  $350 \leq \langle \beta $ seems to be empty of real-world transistors, except possibly for Darlington configurations.  This region of high average current gain is also characterized by large values of 
$s$, as a consequence of the nearly ``concentric'' geometrical organization of the current gain isolines, reaching a peak (in this diagram) of gain,  $V_a$ magnitude and $s$ at the point $(V_a = -130V, s =10 V^{-1})$.

\section{Relationship between Switching Times and Early Parameters}

\begin{figure}[h!]
\centering{
\includegraphics[width=9cm]{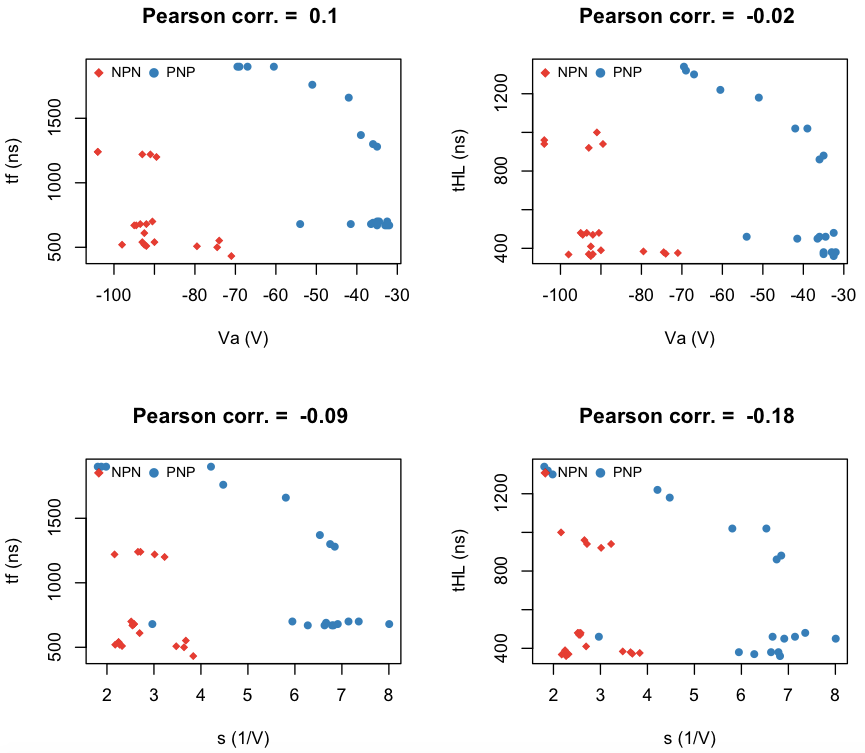}
\caption{Scatterplots of delay and transition times in terms of Early parameters: (a) $t_f \times V_a$; (b) $t_{HL} \times V_a$;.
(c) $t_f \times s$; and ) $t_{HL} \times s$.  No signification relationship can be identified between these pairs of measurements.}
\label{fig:correls}}
\end{figure}

Having experimentally estimated the four switching times and the respective Early parameters for each of the 40
transistors, we are now in position to study the interrelationship between these characteristics.  Figure~\ref{fig:correls}
shows four of the possible correlations between the 4 time measurements and the 2 Early parameters.  Remarkably,
no significant correlation can be identified.  This suggests complete absence of relationships between the considered
delay and transition time measurements and the Early parameters, at least as long as these measurements are
considered isolately.

\begin{figure}[h!]
\centering{
\includegraphics[width=9cm]{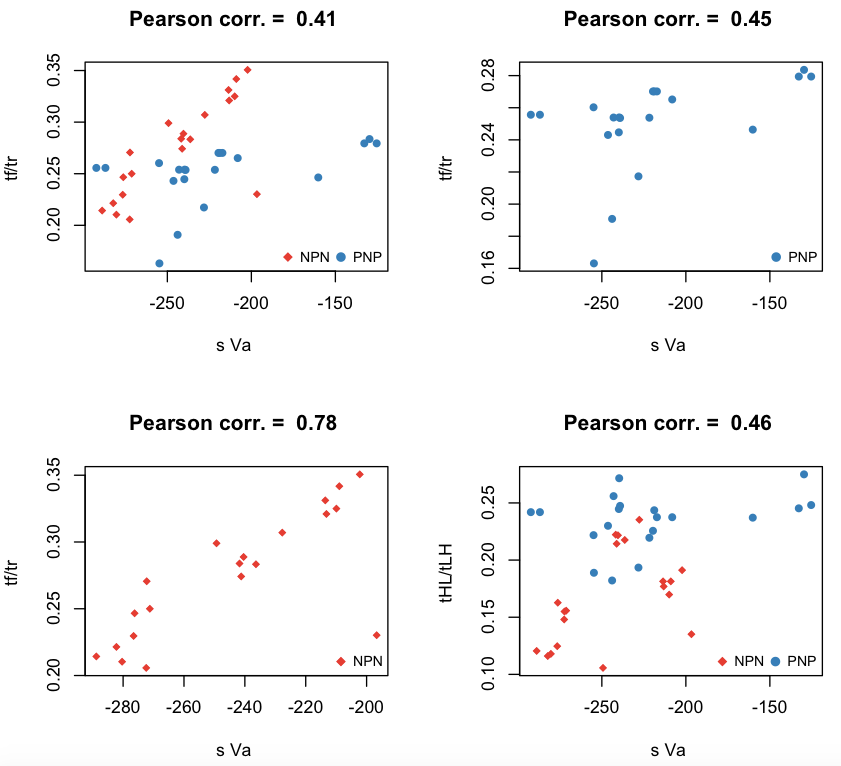}
\caption{Scatterplots and Pearson correlations obtained for: (a) $s V_a \times t_f/t_r$ considering all 40 transistors,
(b) the same, but only PNP devices; (d) the same, for NPN transistors only; and (d)  $s V_a \approx - \langle  \beta \rangle$.
A strong relationship has been observed in (b) with respect to NPN transistors, suggesting that the ratio $t_f/t_r$} can
be, in the average, estimated with good accuracy from the product of Early parameters $s V_a \approx -\langle \beta \rangle$.
\label{fig:ratios}}
\end{figure}

However, it is also interesting to consider combinations between the adopted measurements, such as respective products and
ratios. Figure~\ref{fig:ratios}(a) depicts the scatterplot of $s V_a \times t_f/t_r$. Recall that $\langle \beta \rangle \approx - s V_a$.
Interestingly, a moderate positive correlation
(0.41) is obtained in this case.  However, it can be discerned in this same scatterplot that the NPN and PNP devices follow
different relationships, with the former exhibiting a more definite positive correlation.  Figures~\ref{fig:ratios}(b) and (c) shows
the scatterplot $s V_a \times t_f/t_r$ separately only for PNP (b) and NPN (c) transistors, which yielded respective Pearson
correlations of $0.45$ and $0.78$.  So, the NPN indeed leads to a much more definite relationship $s V_a \times t_f/t_r$, with
a very significant correlation coefficient.  This means that the ratio $t_f/t_r$ can be estimated with relatively good accuracy
from the product $s V_a \approx - \langle \beta \rangle$.  The larger this product, the larger $t_f$ will be relatively to $t_r$.
Figure~\ref{fig:ratios}(d) shows the relationship between $s V_a$ and $t_{HL}/t_{LH}$, which is also characterized by a
moderate positive Pearson correlation of $0.46$.  

So, though no direct relationship can be identified between the four delay and transition times measurements and the two
Early parameters taken isolately,  a relatively strong relationship ultimately appeared when considering the product $s V_a$ and the ratio
$t_f / t_r$.    The obtained positive correlation indicates that more
symmetrical transitions $t_f$ and $t_r$ are obtained for larger values of $s V_a \approx - \langle \beta \rangle$.  This makes
sense because larger current gains will contributed to charging the stray capacitances faster, therefore reducing $t_r$ 
relatively to $t_f$ and yielding larger ratios $t_f/t_r$.  Interestingly, the average current gains typically observed for silicon
NPN and PNP devices (i.e.~$150 \leq \langle \beta \rangle \leq 350$) can lead to a maximum ratio $t_f/t_r$ of about $0.35$.
The symmetric configuration between the rise and fall times, characterized by $t_f/t_r = 1$ cannot be obtained even for the
transistors with peak $\langle \beta \rangle \approx 350$ magnitude, which limits $t_f/t_r$ to $\approx 0.35$.  Thus, NPN
and PNP small signal transistors when used in the adopted circuit configuration (and also likely in other related circuits)
have intrinsic ratios $t_f/t_r$ that are substantially smaller than 1, being therefore characterized
by fall times much smaller than rise times, at least for the considered devices and circuit configuration.

As both Early parameters $V_a$ and $s$ do not reflect any property of the
base-emitter transistor junction, it follows that the observed relationship is mostly concerned with capacitances in parallel with
$R_o = 1/tan(s I_B)$.  The fact that this relationship was almost twice larger for NPN than PNP transistors suggests 
some intrinsic difference regarding the switching properties of these two types of transistors.

\section{Concluding Remarks}

The switching properties of transistors, may they be related to propagation delays or level transitions, critically define the
speed of digital switching circuits.  In the present work, we applied a recently Early modeling approach to 
transistors~\cite{costaearly:2017,costaearly:2018,germanium:2018,costaequiv:2018},
characterized by great simplicity allied to accuracy in representing the transistor non-linearities, as a possible means to study 
switching properties of small signal NPN and PNP silicon bipolar junction transistors. 

Four time measurements were considered for characterizing the transistor switching properties: rise time $t_r$, fall time $t_f$,
transition from low to high levels $t_{LH}$, and transition from high to low levels $t_{HL}$.  The transistor amplifying characteristics
and intrinsic non-linearities stemming from the converging characteristic isolines were effectively summarized in terms of the Early
parameters $V_a$ and $s$.  The time measurements were determined by using a digital oscilloscope, while the Early parameters were obtained by an accurate experimental-numeric, approach previously described~\cite{costaearly:2018} that uses Hough transform 
accumulation in order to identify the intersection of the base current-indexed isoline characteristics implied by the Early effect.

The switching time measurements were found to correlate strongly one another, especially in those cases involving $_f$.  
The Early characterization was in full agreement with results previously obtained with respect to NPN and PNP small signal
silicon transistors~\cite{costaearly:2018}, defining two respective clusters in the Early space $(V_a,s)$.  NPN transistors are characterized by
larger $V_a$ magnitude and smaller $s$ parameter values than PNP transistors.  The average current gain obtained for these
two groups are both very similar to 250.  

The 4 considered switching times exhibited substantial positive Pearson correlation, with the cases involving $t_f$ resulting
in particularly strong correlation values.  When the switching times were individually compared with the Early parameters
$V_a$ and $s$, no appreciable Pearson correlation could be identified.  However, a significant positive Pearson correlation  
was observed between the variables $s V_a \approx - \langle \beta \rangle$ and the ratio $t_f/t_r$.   This correlation resulted much
stronger (0.78) for NPN transistors than their PNP counterparts (0.45), suggesting an intrinsic difference in the relationship between
the switching and Early properties of these two types of small signal silicon transistors.   It is interesting to observe that the
$\langle \beta rangle \rangle$ valued estimated as $-s V_a$ does not necessarily coincides with the $\beta$ values found in data sheets
or obtained by using more traditional approaches.   Indeed, $\langle \beta rangle \rangle$ results from the accurate estimation of
$V_a$ by identification of the isolines characteristics intersection, while $s$ is obtained by linear regression of the isoline angles
$\theta$ and the base current $I_B$.

Interestingly, the ratio $t_f/t_r$ is significantly smaller than 1 for both NPN and PNP types, implying an inherent switching asymmetry.
This asymmetry is reduced for transistors with larger $s V_a \approx - \langle \beta \rangle$, but the maximum ratio obtained for the
considered devices was limited to $t_f/t_r \approx 0.35$.  The fact that NPN and PNP silicon transistors seem to have 
$150 \leq \langle \beta \rangle \leq 350$ implies it to be in principle impossible, at least for the considered devices and circuit configuration,
to reach perfect symmetry between the rise and fall times.  A similar, even though less defined situation holds concerning the
ratio $t_{LH}/t_{HL}$.  Interestingly, in this case no difference resulted regarding the relationship between the times and Early parameters
with respect to the NPN and PNP transistors.

All in all, the results obtained respectively to the considered transistor types and circuit configuration revealed several interesting
results and trends.  In particular, it was shown that the ratio $t_f/t_r$ can be predicted with relatively good accuracy from the
product of Early parameters $s V_a$ of the respective transistor, especially in the case of NPN devices.  This approximation 
can contribute to both practical and theoretical studies of transistors used as switches.  

Several possibilites of further investigations have been defined by the reported approach and respective results, including the
extension to other types of transistors and circuits, Darlington configurations, and power switching.  It would also be
interesting to explain the observed relationships and effects in more theoretical terms.

\vspace{0.7cm}
\textbf{Acknowledgments.}

Luciano da F. Costa
thanks CNPq (grant no.~307333/2013-2) for sponsorship. This work has benefited from
FAPESP grants 11/50761-2 and 2015/22308-2.
\vspace{1cm}

\bibliography{mybib}
\bibliographystyle{unsrt}
\end{document}